\documentclass[preprint]{aastex}

\slugcomment{Accepted for ApJ Letters}
\shorttitle{Cool White Dwarf}
\shortauthors{Harris et al.}

\begin{document}

\title{A New Very Cool White Dwarf Discovered by the Sloan
Digital Sky Survey}

\author{
Hugh C. Harris\altaffilmark{1},
Brad M. S. Hansen\altaffilmark{2},
James Liebert\altaffilmark{3},
Daniel E. Vanden Berk\altaffilmark{4},
Scott F. Anderson\altaffilmark{5},
G. R. Knapp\altaffilmark{2},
Xiaohui Fan\altaffilmark{2,6},
Bruce Margon\altaffilmark{5},
Jeffrey A. Munn\altaffilmark{1},
R. C. Nichol\altaffilmark{7},
Jeffrey R. Pier\altaffilmark{1},
Donald P. Schneider\altaffilmark{8},
J. Allyn Smith\altaffilmark{9},
D. E. Winget\altaffilmark{10},
Donald G. York\altaffilmark{11,12},
John E. Anderson, Jr.\altaffilmark{4},
J. Brinkmann\altaffilmark{13},
Scott Burles\altaffilmark{4,11},
Bing Chen\altaffilmark{14},
A. J. Connolly\altaffilmark{15},
Istv\'an Csabai\altaffilmark{14,16},
Joshua A. Frieman\altaffilmark{4,11},
James E. Gunn\altaffilmark{2},
G. S. Hennessy\altaffilmark{17},
Robert B. Hindsley\altaffilmark{18},
\v{Z}eljko Ivezi\'{c}\altaffilmark{2},
Stephen Kent\altaffilmark{4,11},
D.Q. Lamb\altaffilmark{11},
Robert H. Lupton\altaffilmark{2},
Heidi Jo Newberg\altaffilmark{19},
David J. Schlegel\altaffilmark{2},
Stephen Smee\altaffilmark{14,20},
Michael A. Strauss\altaffilmark{2},
Anirudda R. Thakar\altaffilmark{14},
Alan Uomoto\altaffilmark{14},
Brian Yanny\altaffilmark{4}
}

\altaffiltext{1}{U.S. Naval Observatory, Flagstaff Station, P.O. Box 1149, Flagstaff, AZ  86002-1149}
\altaffiltext{2}{Princeton University Observatory, Princeton, NJ 08544}
\altaffiltext{3}{Steward Observatory, University of Arizona, Tucson, AZ 85721}
\altaffiltext{4}{Fermi National Accelerator Laboratory, P.O. Box 500, Batavia, IL 60510}
\altaffiltext{5}{University of Washington, Department of Astronomy, Box 351580, Seattle, WA 98195}
\altaffiltext{6}{Institute for Advanced Study, Olden Lane, Princeton, NJ 08540}
\altaffiltext{7}{Dept. of Physics, Carnegie Mellon University, 5000 Forbes Ave., Pittsburgh, PA-15232}
\altaffiltext{8}{Department of Astronomy and Astrophysics, The Pennsylvania State University, University Park, PA 16802}
\altaffiltext{9}{University of Michigan, Department of Physics,	500 East University, Ann Arbor, MI 48109}
\altaffiltext{10}{Dept. of Astronomy, University of Texas, Austin, TX 78712}
\altaffiltext{11}{The University of Chicago, Department of Astronomy and Astrophysics, 5640 S. Ellis Ave., Chicago, IL 60637}
\altaffiltext{12}{The University of Chicago, Enrico Fermi Institute, 5640 S. Ellis Ave., Chicago, IL 60637}
\altaffiltext{13}{Apache Point Observatory, P.O. Box 59, Sunspot, NM 88349-0059}
\altaffiltext{14}{Department of Physics and Astronomy, The Johns Hopkins University, 3701 San Martin Drive, Baltimore, MD 21218}
\altaffiltext{15}{Department of Physics and Astronomy, University of Pittsburgh, Pittsburgh, PA 15260}
\altaffiltext{16}{Department of Physics of Complex Systems, E\"otv\"os University, P\'azm\'any P\'eter s\'et\'any 1}
\altaffiltext{17}{U.S. Naval Observatory, 3450 Massachusetts Ave., NW, Washington, DC  20392-5420}
\altaffiltext{18}{Remote Sensing Division, Code 7215, Naval Research Laboratory, 4555 Overlook Ave. SW, Washington, DC 20375}
\altaffiltext{19}{Physics Department, Rensselaer Polytechnic Institute, SC1C25, Troy, NY 12180}
\altaffiltext{20}{Department of Astronomy, University of Maryland, College Park, MD 20742-2421}

\begin{abstract}
Early data taken during commissioning of the SDSS have resulted in the
discovery of a very cool white dwarf.  It appears to have stronger
collision induced absorption from molecular hydrogen than any other
known white dwarf, suggesting it has a cooler temperature than any other.
While its distance is presently unknown, it has a surprisingly small
proper motion, making it unlikely to be a halo star.  An analysis
of white dwarf cooling times suggests that this object may be a low-mass
star with a helium core.  The SDSS imaging and spectroscopy also
recovered LHS~3250, the coolest previously known white dwarf,
indicating that the SDSS will be an effective tool for identifying
these extreme objects.
\end{abstract}


\section{INTRODUCTION}

The identification and study of very cool white dwarf (WD) stars
is important for understanding the evolution of WDs, their luminosity
function, their space density, and their total formation rate.
This study is particularly important for WDs in the Galactic halo,
because they are sufficiently old that most are expected to have
cooled to very low temperatures \citep{han98, ise98, cha99}.
Until recently, the coolest WDs known had temperatures $\sim$4000~K
(Bergeron et al.\ 1997;  Leggett et al.\ 1998).
The discovery \citep{har99, hod00} of two WDs that definitely have
temperatures below 4000~K has shown that such cool WDs do exist and
has added impetus for further searches.

Very cool WDs will have molecular hydrogen in their atmospheres
if any hydrogen is present.  In these high-density atmospheres,
collisions can induce a temporary electric dipole moment in the
hydrogen molecules.  Calculations predict
(Borosow et al. 1989; Lenzuini et al. 1991; Saumon et al. 1994;
Borosow et al. 1997)
a high opacity at infrared wavelengths, referred to as
collision-induced absorption (CIA).  This opacity produces
a spectral energy distribution with greatly depressed
red/infrared flux that should result in distinctive colors
(Bergeron et al.\ 1995a;  Bergeron et al.\ 1997;  Hansen \& Phinney
1998;  Hansen 1998;  Hansen 1999;  Saumon \& Jacobson 1999;
Chabrier 1999).
The Sloan Digital Sky Survey (SDSS) \citep{yor00}
should be especially effective at identifying such stars \citep{har99,
han00}.  This survey is obtaining deep CCD images of 10,000 deg$^2$
of the north Galactic cap in five photometric bands ($u^{\prime}$,
$g^{\prime}$, $r^{\prime}$, $i^{\prime}$, and $z^{\prime}$)
to a limiting magnitude $r^{\prime} \sim 23$ -- see Fukugita et
al. (1996) and Appendix A of Fan et al. (2001) for details of the
photometric system. 

This paper describes the first such star found by SDSS in its
commissioning data.  This star appears to have stronger CIA and,
therefore, is probably cooler than the two previously known sub-4000~K WDs.
In addition, the rediscovery of LHS~3250, previously the WD with
strongest CIA, is reported.  Both (re)discoveries indicate that SDSS
can find other similar stars.

\section{OBSERVATIONS}

Imaging data from the SDSS 2.5 m telescope and its survey camera
\citep{gun98} now include most of the
equatorial stripe with $-1.25^{\rm o} < {\rm{Dec}} < 1.25^{\rm o}$ and outside
of the Galactic plane, as well as a number of scans away from the equator.
The two stars reported here are a new white dwarf, SDSSp J133739.40+000142.8
(hereafter referred to as SDSS 1337+00; ``p'' indicates that the
coordinates are preliminary), and LHS~3250 \citep{luy76}.
They were imaged on 1999 March 21 and 2000 April 04, respectively.
Their positions, proper motions, and magnitudes are given in Table~1\footnote{
The quoted SDSS $asinh$ magnitudes are on the AB system (Lupton et al. 1999),
and errors are internal.
Only preliminary colors and magnitudes are available, until the SDSS
photometric system has been established; $u^*, g^*, r^*, i^*,$ and $z^*$
are used to designate these preliminary values.  Final magnitudes and
colors are not likely to change by much more than a few hundredths.
Note, the difference between $asinh$ magnitudes and traditional logarithmic
magnitudes is negligible at these magnitudes well above the survey limits.}.
A finding chart for SDSS 1337+00 is shown in Figure~1.  The proper motion
is derived from the SDSS position and 
the position in the USNO-A2.0 catalog \citep{mon98} measured on the
Palomar Observatory Sky Survey O and E plates taken in 1952.
For comparison, Table~1 also includes data
taken from the literature \citep{ham99, hod00} for the cool WD
WD0346+246.  Magnitudes in the SDSS system and in the
Johnson/Cousins/CIT system are included in Table~1, where some of the
values not observed directly have been transformed from the other
system using predicted relations for normal stars \citep{fuk96}.
These estimated values are probably accurate to better than 0.1 mag.
The $J$ magnitude limit for SDSS 1337+00 was kindly obtained by F.~Vrba
with the USNO 1.55 m telescope and ALADDIN InSb array detector.

The colors of SDSS 1337+00 and LHS~3250 are highly unusual,
as shown in Figure 2.  They are plotted along with a sample of normal
WDs with synthesized SDSS colors \citep{len98}, and with
observed J magnitudes \citep{ber97, leg98}.  The two curves show the
colors of H and He WD model atmospheres \citep{ber95b} with log~$g = 8$
(kindly calculated and made available to us by
Pierre Bergeron).  The deviation of the spectral energy distributions
of these two stars from normal white dwarfs and non-degenerate stars
is dramatic.  SDSS~1337+00 has colors in Figure 2(b) and 2(c)
even more extreme than those for LHS~3250.

Spectroscopic ``plates'' (exposures with fiber configurations covering
the 3-degree field of view) are taken by SDSS after imaging data are
available --- objects (primarily galaxies and QSO candidates) are
selected for observation based on their image morphology and colors.
QSOs with redshift $3 < z < 4$ have $gri$ colors \citep{fan99, fan01}
similar to those predicted for very cool WDs.
Objects with such colors will be selected
as high-redshift QSO candidates, and those brighter than
$i^{\prime} \approx 20$ will be given high priority for spectroscopic
observation as QSO targets.  They also will be selected
as cool-WD candidates, but given lower priority for spectra.
Both SDSS 1337+00 and LHS~3250 were allocated fibers as QSO candidates.

The two SDSS fiber spectrographs\footnote{
See York et al.\ (2000) and
http://www.astro.princeton.edu/PBOOK/spectro/spectro.htm
for a more complete description of this instrument.
A full description is in preparation (Uomoto et al. 2001).}
cover 3800-9200~\AA\ at a spectral resolution of 1800.
The exposure time is 45~min under optimum observing conditions,
or more as necessary to reach a target S/N.
The spectra are extracted and calibrated with an automated software
pipeline (Frieman et al. 2001, in preparation).
The spectra of SDSS 1337+00 and LHS~3250 were taken on 2000 May 07
(plate 0299, 75 min exposure) and 2000 May 28 (plate 0349, 90 min
exposure), respectively.  They have been smoothed to 6~\AA\ resolution,
and are shown in Figure~3.  The spectrum of LHS~3250
has been scaled to match that of SDSS 1337+00 for ease of comparison;
it agrees quite well with the spectrum previously published \citep{har99}.
There are no significant features in the spectrum of either star.
The spectra show that SDSS~1337+00 has relatively less flux at
red-infrared wavelengths, consistent with the colors shown in Figure~2.

\section{ATMOSPHERIC PARAMETERS}

The low flux at red-infrared wavelengths in SDSS~1337+00 appears similar
to that in LHS~3250 but more pronounced.  The most viable explanation
is that we are seeing CIA from molecular hydrogen \citep{har99}.
Less extreme CIA is also seen in WD0346+246 \citep{hod00},
and possibly in LHS~1126 \citep{ber94} and F351-50 \citep{iba00}.
The discovery of CIA in these stars represents a striking
confirmation of the qualitative trends expected from theoretical
model atmospheres \citep{ber95a, han98, han99, sau99}.
Figure~4 compares the spectrum of SDSS~1337+00 with models of
pure hydrogen atmospheres at three cool temperatures.
The discrepancy between the models and the observed spectrum is most likely
the result of missing opacities in the theoretical models, as well as
the likelihood of the admixture of helium into the hydrogen atmosphere
to provide additional H$_2$-He CIA \citep{ber95a}.  Improvements to
both the atmosphere models and the theoretical opacity inputs are underway.
For example, new models with low temperatures and mixed
(helium-dominated) composition \citep{jor00} show the sensitivity
of CIA and the output spectrum to the input assumptions.
However, these new models only explore a few compositions
out of many plausible possibilities;  they still do not give
improved fits to the spectra in Figure~2.  Until further improved
models are available, we can only deduce broad constraints on the
effective temperature. The fact that the H$_2$ CIA is strong
suggests the objects are somewhat cooler than 4000~K. While
temperatures below 3000~K cannot be ruled out, such low temperatures
imply cooling ages for some ranges of mass that exceed reasonable
estimates for the age of the Galaxy (see Section 4),
so a temperature range 3000-4000~K is suggested.

\section{THE NATURE OF SDSS~1337+00 AND LHS~3250}

Much recent attention has been devoted to the existence of halo WDs,
motivated by microlensing \citep{alc97} and optical \citep{iba99}
observations, and a few WDs are known to be halo stars \citep{lie89}.
Are these two very cool WDs part of the halo population?
The proper motion and distance of WD0346+246 give a tangential velocity
of 175~km~s$^{-1}$, indicating that star is indeed a halo object
\citep{ham99}.  However, the proper motion of LHS~3250 is
smaller (v$_{\rm tan}$ = 81~km~s$^{-1}$) and is consistent with
membership in either the disk or the halo;  meanwhile, its luminosity
is not as low as expected for a presumably very old halo WD.
The proper motion of SDSS~1337+00 is smaller yet.  If it has
an absolute magnitude similar to that of LHS~3250 ($M_V$ = 15.7),
then its distance will be about 54~pc and v$_{\rm tan}$ will be
46~km~s$^{-1}$.  Alternatively, if it has the lower luminosity
($M_V \sim 17.5-18$) expected for a conventional, 0.6-$M_{\odot}$ WD
cooling to a temperature where CIA becomes strong,
then its distance will be about 20~pc and v$_{\rm tan}$ will be
only 17~km~s$^{-1}$.  In either case (pending a parallax measurement),
these two coolest WDs do not have the properties expected for halo stars.

To properly constrain the nature of these objects, we
must consider not only WDs with conventional C/O cores, but also
the known population of low mass (${\sim}0.45 M_{\odot}$ or less)
helium core WDs found in binaries \citep{ber92, mar95}.
Such objects are the result of truncated stellar evolution in close
binaries \citep{kip67} and cool more slowly (for similar masses)
as a result of the increased heat capacity of the helium core.
Figure~5 shows the cooling time for WDs with hydrogen atmospheres
(Hansen 1999; see also Benvenuto \& Althaus 1999, Chabrier et al.\ 2000,
Salaris et al.\ 2000).
Conventional WDs ($\sim 0.6 M_{\odot}$) require ages $>$8.5~Gyr
to cool to 4000~K. WDs more massive than average can cool somewhat
faster at late times due to the earlier onset of core crystallization.
The aforementioned low mass helium core WDs can potentially cool below
4000~K much faster. A 0.23 $M_{\odot}$ helium core WD may cool below
4000~K in $\sim$6~Gyr.  (Note that very low mass WDs may possess
thick hydrogen envelopes and thus will cool more slowly, with a
contribution to their luminosity from residual hydrogen burning
\citep{dri98}.  If true, this will cause the curves shown
in Figure~5 to turn upwards again below $\sim 0.25 M_{\odot}$.)

Further progress requires that we
determine a temperature or a radius, because the luminosity is known
(or soon will be, after the parallax of SDSS 1337+00 is measured).
The radius varies from $1.4 \times 10^9$~cm for the $0.23 M_{\odot}$
model, to $5.3 \times 10^8$cm for the $1.0 M_{\odot}$ model, so the
luminosity at fixed temperature varies by a factor of 7
between these two possibilities.  Fitting the spectrum
with models may give very accurate temperatures when the models
are improved (Section 3).
The absolute magnitude of LHS~3250 ($M_V=15.72$)
suggests a low mass, helium core WD as the most likely of the
above options. In many respects SDSS~1337+00 is similar --- the
proper motion and apparent magnitude are both consistent with it being an
LHS~3250 analog, but slightly cooler, at two to three times the distance.
Both stars are more likely to belong to the old disk or thick disk
populations than to the halo.

Most WDs have masses $\sim 0.6 M_{\odot}$. Both low mass and
high mass WDs are minority constituents. Thus it seems
surprising that such potentially anomalous WDs are
among the first cool WDs found in the SDSS survey. This may, in
part, be due to the strong color selection involved. WDs are
only selected for spectroscopic follow-up if they appear to lie outside
the main stellar locus in the $griz$ passbands.
Thus, the accelerated evolution to low temperatures at the high-
and low-mass ends of the distribution may favor their detection.
Identifying cool WDs with temperatures about 3500-4500~K
that do not have such strong CIA is more difficult.
Transition stars like WD0346+246 may be found
using a combination of visible and near-infrared colors, or using an
intermediate-band filter to find stars with no MgH absorption feature
\citep{cla95, met00}, and then doing follow-up spectroscopy.
With spectroscopic plates taken and examined
covering roughly 400 square degrees of sky, we have found two cool
WDs (one previously known).  Thus we might
expect to find at least several tens in the full SDSS survey.

\acknowledgments
We are grateful to F. Vrba for obtaining the J magnitude limit of
SDSS 1337+00.
The Sloan Digital Sky Survey (SDSS) is a joint project of the University
of Chicago, Fermilab, the Institute for Advanced Study, the Japan
Participation Group, the Johns Hopkins University, the
Max-Planck-Institut f\"{u}r Astronomie, New Mexico State University,
Princeton University,
the United States Naval Observatory, and the University of Washington.
Apache Point Observatory, site of the SDSS, is operated by the
Astrophysical Research Consortium.   Funding for the project has been
provided by the Alfred P. Sloan Foundation, the SDSS member institutions,
the National Aeronautics and Space Administration, the National Science
Foundation, the U.S. Department of Energy, Monbusho, and the
Max Planck Society.
The SDSS World Wide Web site is http://www.sdss.org/.
Support for this work was provided by NASA through Hubble Fellowship
grant \#HF-01120.01-99A from the Space Telescope Science Institute,
which is operated by the Association of Universities for Research in
Astronomy, Inc. under NASA contract NAS5-26555.

\clearpage

\figcaption[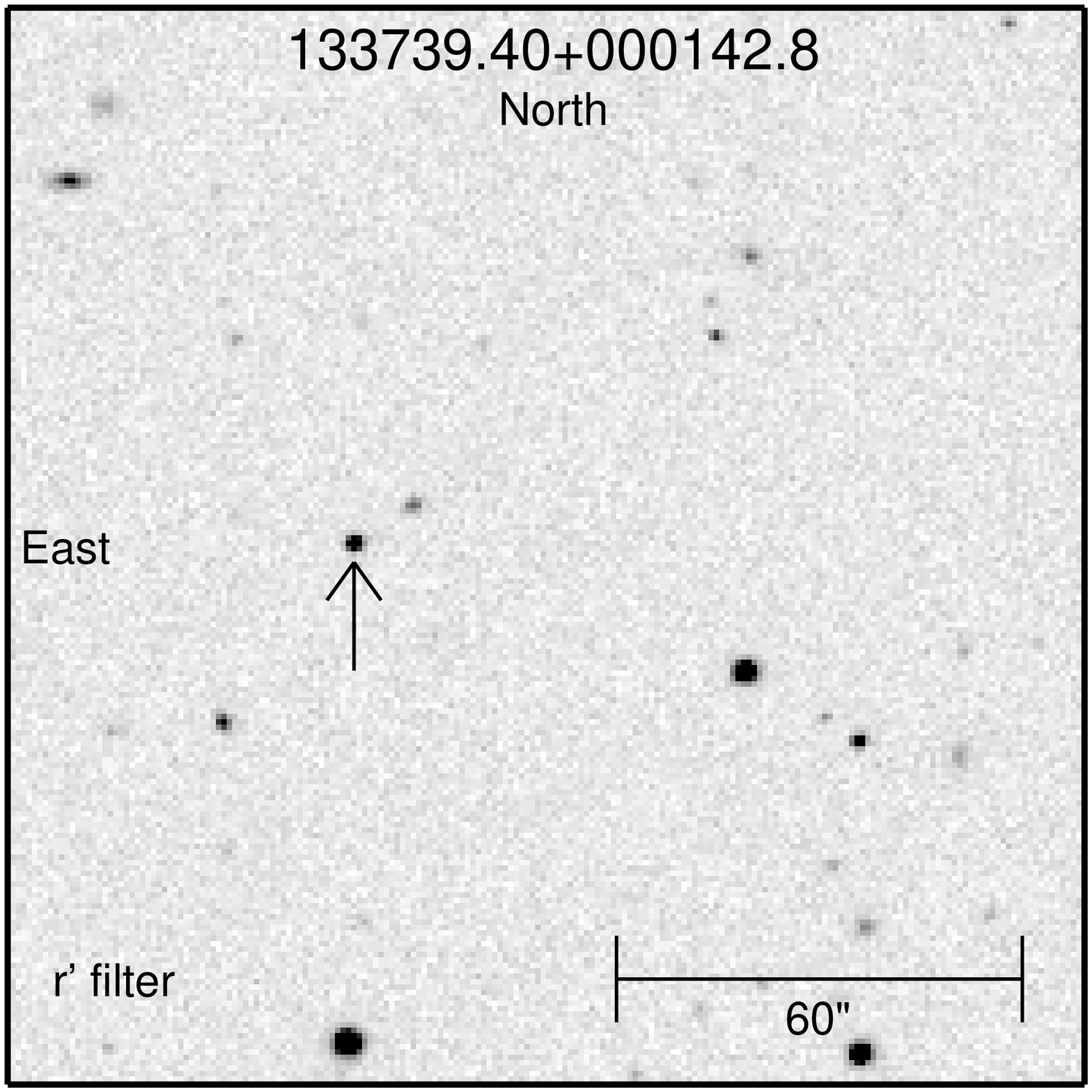]{
Finding chart for SDSSp J133739.40+000142.8 taken from the SDSS
$r'$ image observed on 1999 Mar 21.  The frame is 160$''$ on a side.}

\figcaption[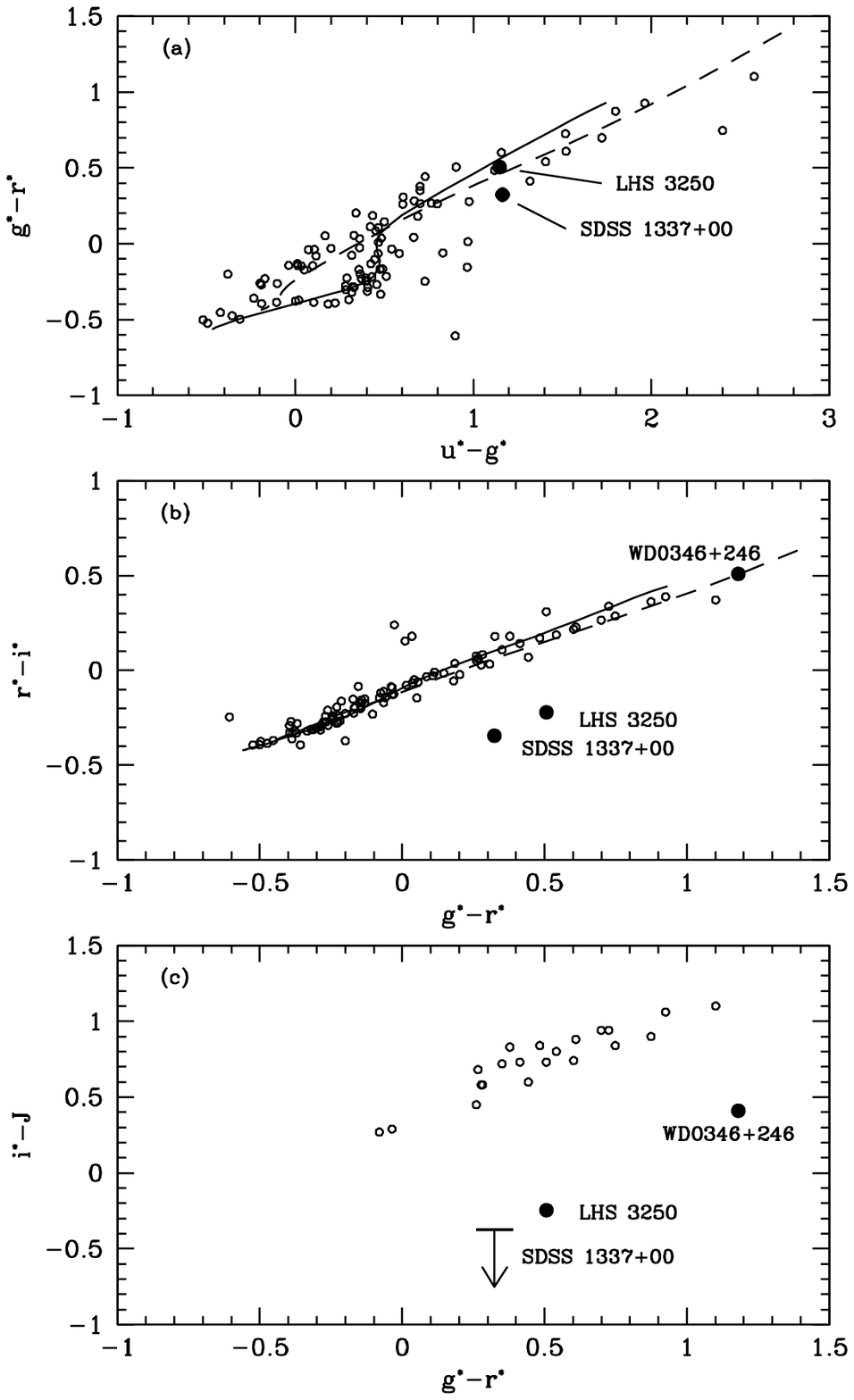]{
Three color-color diagrams showing a sample of normal WDs and the
two very cool stars reported here.
The curves show the colors of model
atmospheres \citep{ber95b} of pure H (solid curves) and pure He
(dashed curves) with log~$g = 8$.}

\figcaption[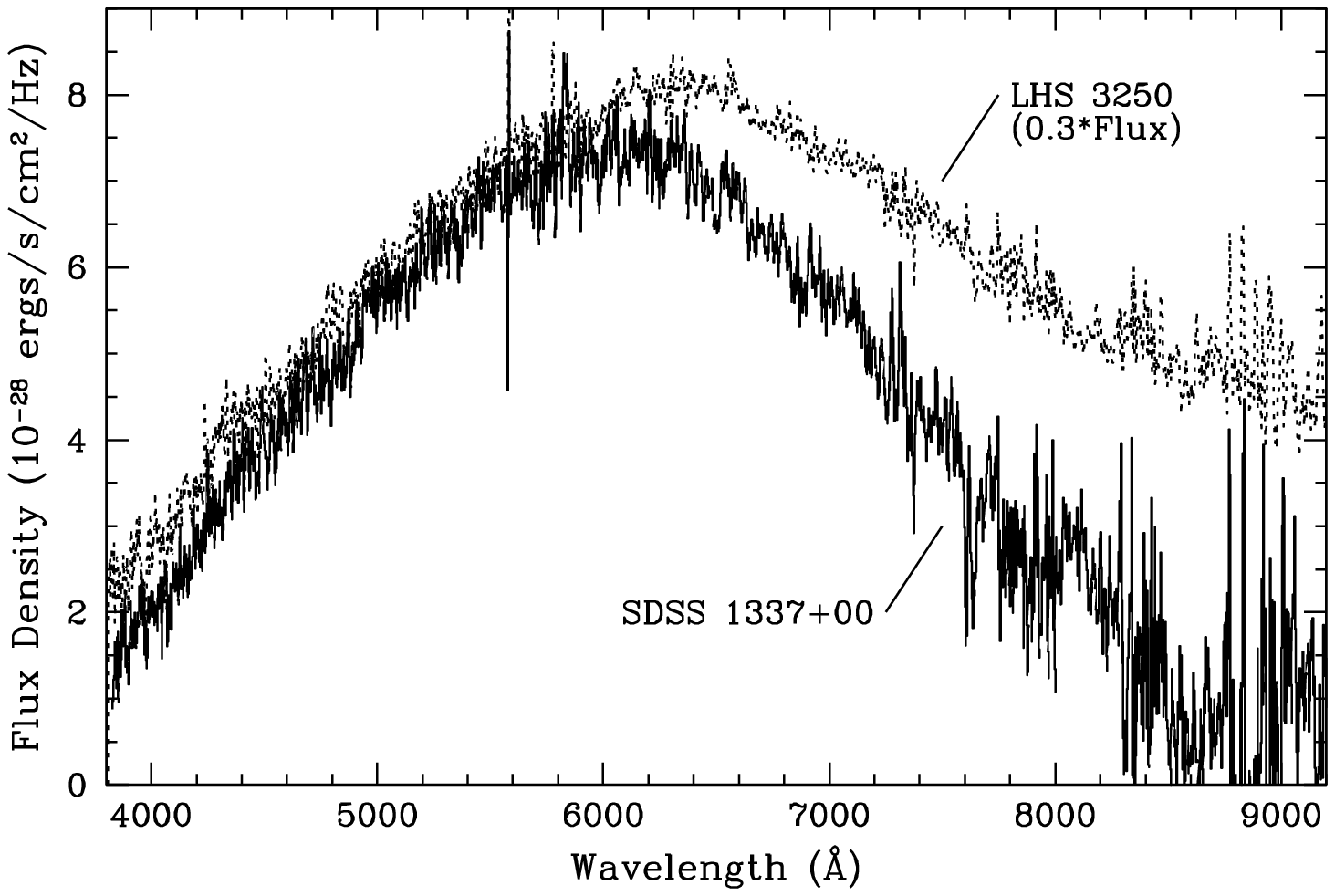]{
Spectra taken with the SDSS 2.5-m telescope multifiber spectrograph
of the known cool WD LHS~3250 and the new, more extreme star SDSS 1337+00.
The flux of LHS~3250 has been scaled by a factor of 0.3
to show its close match to SDSS 1337+00 at the blue end of the spectrum.}

\figcaption[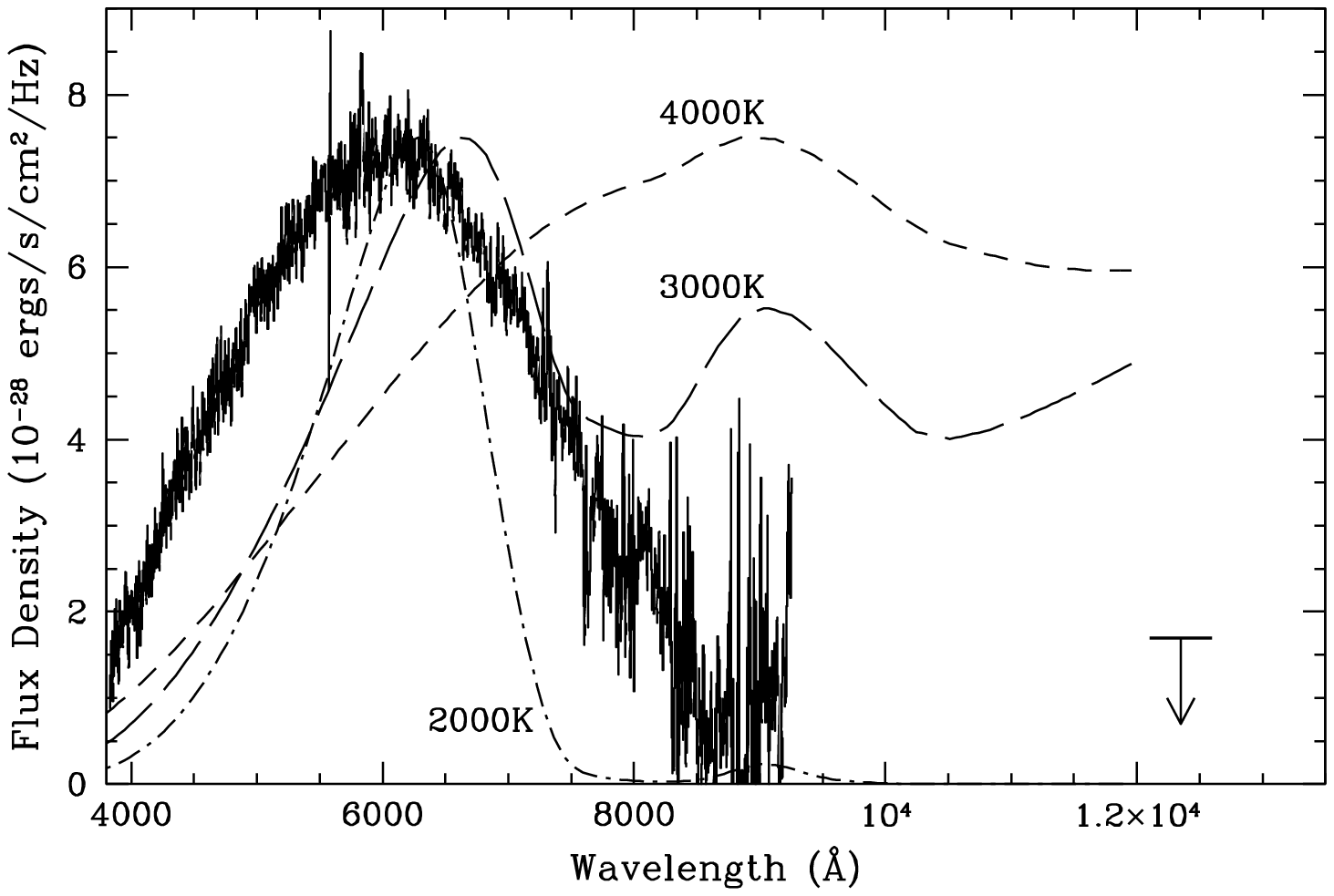]{
Spectrum of SDSS 1337+00 compared with preliminary models of cool,
pure hydrogen atmospheres.  None of the models in this temperature
sequence adequately fits the data (see text).}

\figcaption[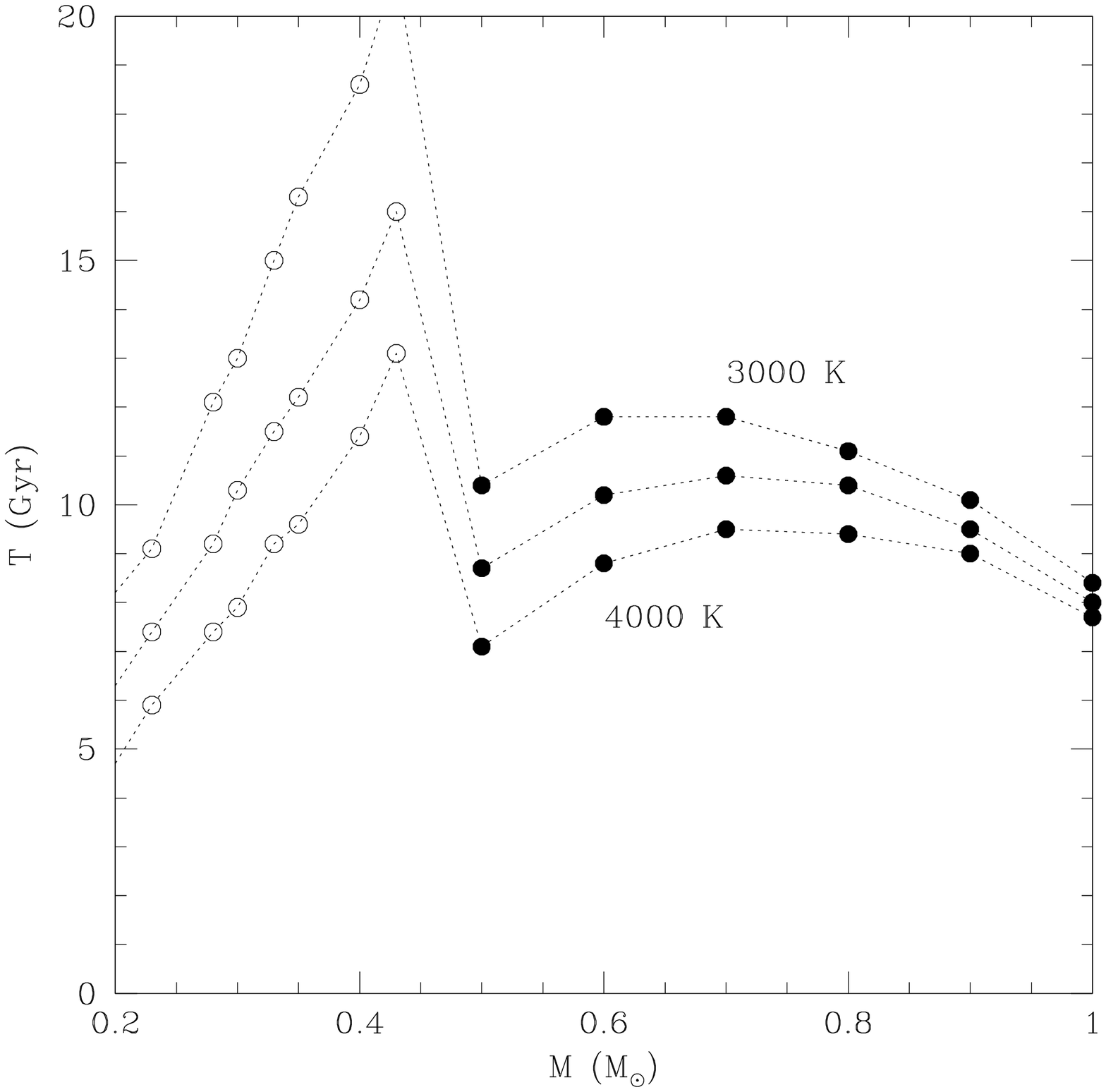]{
Cooling times for pure hydrogen models of a given mass to reach
a temperature of 4000~K, 3500~K, and 3000~K.
Stars with $M \ge 0.50 M_{\odot}$ and a C/O core are shown by filled circles;
stars with $M \le 0.45 M_{\odot}$ and a He core are shown by open  circles.}


\begin{deluxetable}{llll}
\tablenum{1}
\tablewidth{4.3 in}
\tablecaption{Observational Data}
\tablehead{
\colhead{Parameter} &
\colhead{SDSS 1337+00} &
\colhead{LHS 3250} &
\colhead{WD0346+24\tablenotemark{a}} 
}
\startdata
RA  & 13 37 39.4 & 16 54 01.3 & 03 46 46.5 \\
Dec & 00 01 43 & 62 53 55 & 24 56 04 \\
$\mu$ (mas yr$^{-1}$) & 182$\pm$10 & 566 & 1302 \\
PA (degrees)         & 185$\pm$3 & 286 & 155 \\
$u^*$ & 20.67$\pm$0.06 &19.52$\pm$0.04 &  \dots \\
$g^*$ & 19.50$\pm$0.01 &18.37$\pm$0.02 &  19.7\tablenotemark{d}  \\
$r^*$ & 19.18$\pm$0.01 &17.87$\pm$0.02 &  18.5\tablenotemark{d}  \\
$i^*$ & 19.53$\pm$0.02 &18.09$\pm$0.01 &  18.0\tablenotemark{d}  \\
$z^*$ & 19.98$\pm$0.09 &18.55$\pm$0.04 &  \dots \\
$B$   & 19.8\tablenotemark{d} &18.85$\pm$0.02 & $>$20.4 \\
$V$   & 19.3\tablenotemark{d} &18.07$\pm$0.01 &  19.06$\pm$0.01 \\
$R$\tablenotemark{b} & 19.1\tablenotemark{d} &17.74$\pm$0.02 & 18.30$\pm$0.06\\
$I$\tablenotemark{b} & 19.2\tablenotemark{d} &17.87$\pm$0.02 & 17.54$\pm$0.02\\
$J$\tablenotemark{c} &$>$19.9           &18.33$\pm$0.05 & 17.60$\pm$0.05 \\
\enddata
\tablecomments{Units of right ascension are hours, minutes,
and seconds, and units of declination are degrees, arcminutes, and
arcseconds.  Coordinates are given for equinox and epoch 2000.}
\tablenotetext{a}{Values taken from Hodgkin et al. (2000) and
  Hambly et al. (1999).}
\tablenotetext{b}{$R$ and $I$ magnitudes are on the Cousins system.}
\tablenotetext{c}{$J$ magnitudes are on the CIT system.}
\tablenotetext{d}{Values estimated from transformations in
  Fukugita et al. (1996).}
\end{deluxetable}

\plotone{hchwdfig1.eps}
\clearpage
\plotone{hchwdfig2.eps}
\clearpage
\plotone{hchwdfig3.eps}
\clearpage
\plotone{hchwdfig4.eps}
\clearpage
\plotone{hchwdfig5.eps}
\end{document}